\documentclass[%
reprint,
amsmath,amssymb,
aps,
pra,
]{revtex4-2}
\usepackage{booktabs}
\usepackage{graphicx}
\usepackage{dcolumn}
\usepackage{bm}
\usepackage{lipsum}
\usepackage{multirow}
\usepackage{xcolor}
\usepackage{relsize}
\usepackage{siunitx}
\usepackage{xparse}
\usepackage{natbib}
\usepackage{textgreek}
\usepackage[caption=false]{subfig}

\begin{document}
\preprint{APS/123-QED}

\title{Full molecular dynamics simulations of a single trapped ion in a neutral bath}
\author{Saajid Chowdhury}
\author{Ruiren Shi}
\author{Jes\'us Pérez-Ríos}
\affiliation{Department of Physics and Astronomy, Stony Brook University, Stony Brook, NY 11794, USA}

\begin{abstract}
We present full molecular dynamics simulations that explicitly incorporate the simultaneous interactions between a trapped ion and a bath of neutral atoms. In contrast to conventional molecular dynamics treatments, this framework enables a systematic assessment of how the atomic-gas density influences both the ion’s cooling dynamics and its steady-state mean kinetic energy. Our results show that the gas density measurably modifies the ion’s average kinetic energy, albeit only weakly, in qualitative disagreement with predictions obtained from standard molecular dynamics simulations. In addition, the calculations indicate that short-range features of the atom–ion interaction potential become increasingly consequential as the atomic density increases. Simulations including many atoms yield higher mean kinetic energies than standard molecular dynamics, a trend we attribute to the transient formation of molecular-ion complexes. Finally, we observe that for a fixed ion, lighter atomic baths lead to a smaller final ion average kinetic energy than heavier atomic baths.
\end{abstract}

\maketitle

\section{Introduction}

Hybrid ion–atom systems combine trapped ions with ultracold atomic gases, providing a platform for investigating interactions between charged and neutral particles~\cite{Tomza2019,LOUS202265,COTE20166,Perez-Rios2020,COTE20166,Deiss2024}. Studies of cold collisions and chemical reactions have revealed the rich dynamics arising from these interactions~\cite{Hall_2011,Hall_2012,Willitsch_2008,Toscano_2020,Heazlewood18072019,Harter2014,Krukow2016,JPR_2021}, while the observation of ion–atom Feshbach resonances has opened a route to controlling them~\cite{Weckesser2021,Schaetz_2025}, analogous to the control achieved in ultracold neutral gases~\cite{Chin_RMP}. Realizing this potential requires control over the ion’s translational motion, which is governed by its interactions with both the atomic bath and the trapping field. In a radiofrequency trap, collisions can couple the driven micromotion to the ion’s secular motion, producing a nonequilibrium steady state whose energy distribution differs from that of the surrounding gas~\cite{Rouse_2017,Zipkes_2011}. Understanding this energy exchange is therefore essential for controlling collision energies and accessing the quantum regime of ion–atom interactions.

The energy relaxation of a single trapped ion immersed in an atomic gas has been investigated using several complementary approaches. Within classical descriptions, these include Monte Carlo collision models~\cite{Zipkes_2011}, Langevin-type stochastic equations~\cite{Londono_2022}, and molecular dynamics (MD) simulations~\cite{Pinkas2020,furst2018prospects}. Monte Carlo collision models propagate the ion in the trapping field between instantaneous collisions, whose occurrence and outcomes are sampled according to prescribed collision rates and scattering laws, often based on the Langevin capture model. Langevin-type approaches instead represent the bath through friction and fluctuating forces, with the ion–atom interaction entering the associated transport coefficients. MD simulations resolve the collision by propagating the coupled motion of the ion and a single atom under their interaction potential and the trapping field. These simulations capture finite-duration encounters, collision-induced energy exchange with the trap, and the formation of transient ion–atom complexes.

Despite their detailed treatment of individual encounters, commonly used MD implementations introduce atoms sequentially, allowing only one atom to interact with the ion at a time. We refer to this implementation as the “one-atom-at-a-time” approach. This approximation describes isolated binary encounters but excludes the arrival of an additional atom while a collision or transient complex is still evolving. Its validity therefore depends on the separation between the duration of an encounter and the waiting time for another atom to reach the interaction region. Long-lived ion–atom complexes can challenge this separation even in a dilute gas, allowing additional atoms to participate before the complex dissociates~\cite{Hirzler2023,Pinkas2023}. In particular, the one-atom-at-a-time approach cannot capture stabilization of a transient complex through a collision with a third particle, a mechanism relevant to the formation of weakly bound ionic complexes in cold buffer gases, including helium-tagged ions~\cite{BRUNKEN201767}. As the bath density increases, additional encounters become more likely, raising the question of how complex formation and stabilization affect the ion’s energy relaxation.

In this work, we develop a full molecular dynamics  simulation of a single trapped ion interacting with an atomic bath, allowing multiple atoms to participate in the dynamics simultaneously. This framework enables us to investigate how overlapping encounters and transient complexes affect energy equilibration as the bath density varies. We find that the ion’s energy relaxation depends on both bath density and the short-range form of the ion–atom potential, though only slightly, and that this behavior is not reproduced by the one-atom-at-a-time implementation considered here. We further identify a limitation in the sampling of incoming atoms in this implementation: neglecting the ion’s motion when constructing the incident flux can misrepresent the distribution of positions and relative velocities for the sampled encounters. Our results delineate the conditions under which sequential binary-collision models describe ion cooling and identify regimes that require an explicit treatment of the many-atom bath.

\begin{figure}
    \includegraphics[width=\linewidth]{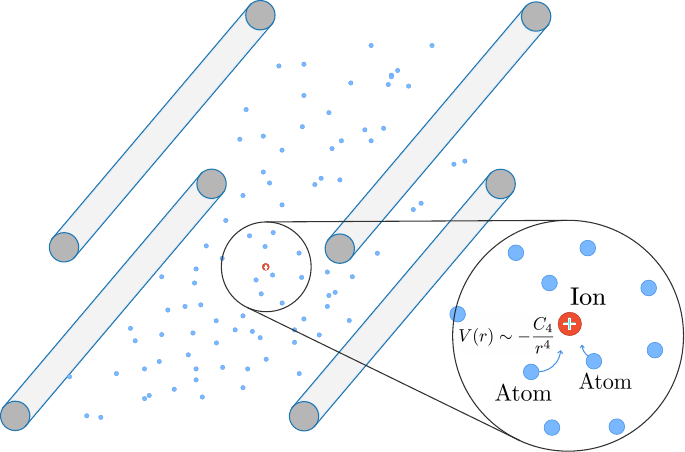}
    \caption{A schematic of our system, which shows the four electrodes of the Paul trap, the trapped ion in red, the ultracold atomic cloud surrounding the ion, an inset showing 10 atoms within a sphere surrounding the ion, and the atom-ion interaction, which is a charge-induced-dipole interaction $V(r)\sim -C_4/r^4$ in the long range.}
    \label{fig:fig1}
\end{figure}

\section{One-atom-at-a-time approach}

Our first simulation method is the one-atom-at-a-time approach, closely resembling the conventional method for simulating a trapped ion in a low-density gas \cite{furst2018prospects}. In this approach, the ion starts at rest at the center of the Paul trap and is treated as a point particle which is simulated classically in the Paul trap's time-dependent electric field. Given an interaction sphere radius, according to the atomic density, the ion interacts with only one atom at any given time during the simulation. After each atom is done interacting with the ion, the atom is removed, and then another atom is spawned in. The trapped ion gradually gets more kinetic energy, so it gets "hotter", as it encounters more atoms and strays further from the trap center. This method assumes that the ion interacts with only one atom at a time because the typical atom density, $10^{11}$-$10^{13}$ cm$^{-3}$, implies that within a large sphere of radius approximately 300-1300 nm surrounding the ion, there is typically only one atom. 

We use several model potentials $V(r)$ for the atom-ion interaction, including a Lennard-Jones type potential $V_8(r)=C_8/r^8-C_4/r^4$, which has a long-range ion-induced-dipole tail of $-C_4/r^4$ where $C_4=\alpha/2$ and $\alpha$ is the polarizability of the atom in atomic units. Including the trap, the equations of motion for the ion's position $\vec{r}_i$ and the atom's position $\vec{r}_a$ are
\begin{align}
    \ddot{r}_{i,j}&=-\frac{1}{m_i}\frac{\partial V(r_{ai})}{\partial r_{i,j}}-\frac{\Omega_\textrm{rf}^2}{4}(a_j+2q_j\cos(\Omega_\textrm{rf}t))r_{i,j}\\
    \ddot{r}_{a,j}&=-\frac{1}{m_a}\frac{\partial V(r_{ai})}{\partial r_{a,j}},
\end{align}
where $j=x,y,z$, $r_{ai}=|\vec{r}_{a}-\vec{r}_i|$ is the atom-ion distance, $r_{i,j}$ is the $j^\textrm{th}$ coordinate of the ion, and $r_{a,j}$ is the $j^\textrm{th}$ coordinate of the atom. Each atom is spawned in at a uniformly random point on the surface of a sphere of radius $R$ centered at the ion, which is calculated from the number density of the gas as $n=(\frac{4}{3}\pi R^3)^{-1}$. The newly spawned atom's velocity is sampled as described in Ref.~\cite{furst2018prospects}: the inwards radial component $-v_r$ (towards the ion) is sampled from a Weibull distribution, which is a Rayleigh distribution with shape parameter 2 and scale $\sqrt{k_BT_a/m_a}$, and the tangential components $v_\theta$ and $v_\phi$ are independently sampled from Gaussians with standard deviation $\sqrt{k_BT_a/m_a}$, where $T_a$ is the atom bath temperature and $m_a$ is the atom mass. After the atom finishes interacting with the ion for 1-2 ms, the atom is removed and the ion is propagated alone, following the Mathieu equations, for 12-24 $\mu$s (10-20 $x$-secular periods) to track and record the ion's average kinetic energy while it is alone. A new atom is then spawned, and the process repeats thousands of times, yielding thousands of collisions for a single trajectory. From one step to the next, the ion's positions and velocities and the time $t$ are always carried over continuously, preventing discontinuities. The average kinetic energies of the ion during each lone-propagation interval serves as a proxy for the temperature of the ion as a function of time.

\section{Full Molecular Dynamics Simulations}\label{sampling}

To arrive at the correct thermalization rate and time, it is important to correctly sample the positions and velocities of the atoms interacting with the trapped ion. The method in the previous section samples atom positions uniformly around the ion, samples tangential velocity components from Gaussians, and samples the radial velocity component from a Weibull distribution. In this section, we explain a different method of sampling incoming atom positions and velocities that better captures the environment around the ion, accurately accounting for the ion's motion. This different method is used in our full MD simulations.

The trapped ion is immersed in a low-density ($10^{11}$-$10^{13}$ cm$^{-3}$) gas at microkelvin temperatures, so the atoms' speeds have a Maxwell-Boltzmann distribution, and their Cartesian velocity components $v_j$, $j=x,y,z$ independently follow identical Gaussian distributions: 
\begin{equation}\label{eq:boltzmann}
    P(v_j)=\sqrt{\frac{m}{2\pi k_BT}}e^{-mv_j^2/2k_BT}.
\end{equation}
It is not necessary to simulate all of the atoms in the gas, since only the closest few atoms will interact with the ion at any given time. When an atom drifts sufficiently far away, it no longer needs to be simulated, and when an atom approaches close enough to the ion, it should become included in the simulation. 

It turns out that the position and velocity distribution of the subset of atoms that approach the ion is not the same as the position and velocity distributions of all the atoms in the gas. First of all, if the ion is moving very fast in the $+z$ direction, then it will encounter more atoms coming from the $+z$ direction compared to the $-z$ direction. Second, although faster-moving atoms occupy a lower number density in the gas than slower-moving atoms (due to the Boltzmann weight in Eq.~\ref{eq:boltzmann}), the number of atoms per unit time that approach the ion is proportional to both their number density and the component toward the ion of their velocity relative to the ion. To account for this, in our full MD simulations, atoms sampled from the Maxwell-Boltzmann gas are subject to flux-weighted rejection sampling, where the flux takes into account the relative velocity between the proposed atom and the ion, similar to Maxwellian inflow distributions simulated via surface-flux-weighted rejection sampling described elsewhere \cite{garcia2006,bird1994}. These methods are often described for stationary surfaces, but they can be applied to moving surfaces by working in the reference frame co-moving with the surface. 

To be more precise, suppose that there are a very large number $N$ atoms within a large box of width $L$, and suppose that the ion is currently moving at velocity $\bf{u}$. Let $r\ll L$ be the radius of a small sphere centered at the ion's position (the origin of our spherical coordinate system) at all times. The atoms that approach the ion are the atoms that enter this sphere. Let $n$ be the number of these atoms, during time interval $dt$, which enter the moving sphere of radius $r$ at an infinitesimal range of solid angle $\theta\in[\theta,\theta+d\theta]$ and $\phi\in[\phi,\phi+d\phi]$ with velocity ${\bf{v}}\in[{\bf{v}},{\bf{v}}+d{\bf{v}}]$. 

\begin{figure}[h]
    \centering
    \includegraphics[width=\linewidth]{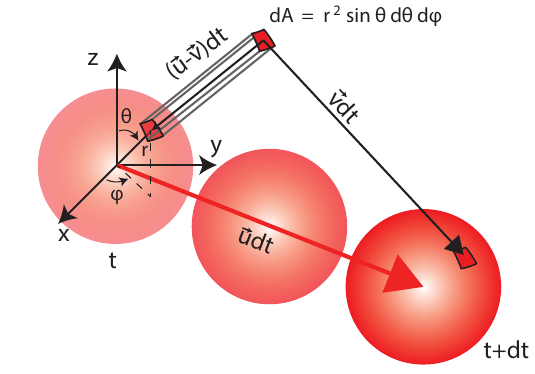}
    \caption{The position of the ion at times $t$ and $t+dt$, the sphere of radius $r$ around the ion at the initial and final time, and the prism of atoms moving at velocity $\bf{v}$ which approach the ion at the infinitesimal rectangular patch of area $r\sin\theta\,d\phi\times r\,d\theta$.}
    \label{fig:fig2}
\end{figure}

In time $dt$, the ion will have moved a distance $u\,dt$, and the atoms with velocity $\approx\bf{v}$ which will approach the ion during the end of the time interval $dt$ will have traveled a distance $v\,dt$. In Fig.~\ref{fig:fig2}, the vector triangle with legs ${\bf u}\,dt$ and ${\bf v}\,dt$ has as its third side $({\bf v}-{\bf u})\,dt$. The set of atoms which approach the ion at $(\theta,\phi)$ during the time interval $dt$ initially occupy a thin prism of space of length $|{\bf v}-{\bf u}|\,dt$. The base of this prism has area $r^2\sin\theta\,d\theta\,d\phi$, which is the infinitesimal piece of spherical surface area which the atoms pass through. The prism of atoms has volume $V=(r^2\sin\theta\,d\theta\,d\phi)(|{\bf v}-{\bf u}|\,dt)|\hat{\bf r}\cdot \hat{\bf v}_\textrm{rel}|$, where ${\bf v}_\textrm{rel}:={\bf v}-{\bf u}$ is the relative velocity of the atoms with respect to the ion, and we project this in the direction of the position on the sphere $\hat{\bf r}:=(\sin\theta\cos\phi,\sin\theta\sin\phi,\cos\theta)$ since the prism is slanted. The total number of atoms in the box that appear within the prism is $NV/L^3$, and the fraction of those atoms with velocity ${\bf v}\in[{\bf v},{\bf v}+d{\bf v}]$ is $\left(\frac{m}{2\pi k_BT}\right)^{3/2}e^{-\frac{m(v_x^2+v_y^2+v_z^2)}{2k_BT}}\,dv_x\,dv_y\,dv_z$. Putting it all together, we have
\begin{equation}
    \begin{aligned}
        n=&\frac{N}{L^3}(r^2\sin\theta\,d\theta\,d\phi)(|{\bf v}-{\bf u}|\,dt)|\hat{\bf r}\cdot \hat{\bf v}_\textrm{rel}|\cdot \\
        &\left(\frac{m}{2\pi k_BT}\right)^{3/2}e^{-\frac{m\left(v_x^2+v_y^2+v_z^2\right)}{2k_BT}}\,d^3{\bf v}.
    \end{aligned}
\end{equation}
Thus, the probability density of positions and velocities of atoms which approach the ion, given that the ion's velocity is ${\bf u}$, is
\begin{equation}
    P({\bf v},\theta,\phi |\, {\bf u})=A\max(0,-v_{\textrm{rel},r})e^{-\frac{mv^2}{2k_BT}}\sin\theta
\end{equation}
where $A$ is a normalization constant and $v_{\textrm{rel},r}$ is the radial component of ${\bf v}_\textrm{rel}={\bf v}-{\bf u}$. We write $\max(0,-v_{\textrm{rel},r})$ instead of just $|v_{\textrm{rel},r}|$, because atoms moving away from the ion in the ion's frame will not approach the ion during time $dt$. When the ion is not moving, $u=0$ and therefore the velocity distribution is equivalent to  $P({\bf{v}})=A\max(0,-v_r)e^{-\frac{mv_r^2}{2k_BT}}e^{-\frac{mv_\theta^2}{2k_BT}}e^{-\frac{mv_\phi^2}{2k_BT}}$, where the radial component towards the ion $-v_r$ follows a Weibull distribution as is the case for the one-atom-at-a-time approach.

\subsection{Computational approach}~\label{comput}

We assume that the ion is continuously interacting with 10 atoms. The interaction between the ion with each atom is included, but the atom-atom interactions are not included \footnote{We find that including atom-atom Lennard-Jones interactions does not have a measurable effect on the ion's average kinetic energy for the Yb$^+$-Rb system at density $10^{14}$ cm$^{-3}$.}. At the start of a trajectory, the atoms spawn uniformly randomly within the volume of a sphere of radius $R$ (where $n=10(\frac{4}{3}\pi R^3)^{-1}$, so this $R$ is larger than the one-atom-at-a-time case) around the ion, with all 3 Cartesian components of velocity sampled from Gaussians with standard deviation $\sqrt{k_BT_a/m_a}$. The ion interacts continuously with these 10 atoms, and whenever one atom drifts farther away from the ion than $R$, that atom is removed and then another atom spawns at a distance $R$ with an inwards radial velocity component relative to the ion. The atom's position $\hat{n}$ and velocity $\vec{v}$ are chosen by rejection-sampling from
\begin{equation}
    P(\vec{v},\hat{n}|\,\vec{u})\propto e^{-mv^2/2k_BT_a}\max(0,-(\vec{v}-\vec{u})\cdot\hat{n}),
\end{equation}
where $\vec{u}$ is the ion's velocity, and $-(\vec{v}-\vec{u})\cdot\hat{n}$ is the inwards radial component of the relative velocity of the atom with respect to the ion. The direction $\hat{n}$ is initially drawn uniformly from all possible directions, and all Cartesian components of $\vec{v}$ are initially drawn from independent Gaussians with standard deviation $\sqrt{k_BT_a/m_a}$, before being filtered via rejection sampling by a flux-weighting proportional to $\max(0,-(\vec{v}-\vec{u})\cdot\hat{n})$, which accounts for the greater rate at which radially-faster atoms see the ion and the greater probability of the ion to encounter atoms coming from the direction towards which the ion is moving. 

The ion's instantaneous kinetic energy is recorded every few simulated microseconds throughout the simulation. For each ensemble, we first run 28-96 trajectories for 8 hours of wall time and analyze the kinetic energy as a function of simulated time to determine an estimate of the thermalization time $T_0$ of the ensemble (typically 5~ms for the highest densities through 1000~ms for the lowest densities). We then estimate how many timesteps it would take to simulate a trajectory up to $2T_0$ of simulated time, assuming it does not form a molecular ion during the trajectory. We then run as many trajectories as possible with this fixed number of timesteps on 96 CPU cores in 8 hours, yielding 1000 to 50000 trajectories, depending on the number of timesteps per trajectory. The ion's average kinetic energy in this ensemble is then the average ion instantaneous kinetic energy over every seeds' recorded CSV rows (evenly spaced in simulated time) after $T_0$ during which the ion wasn't currently bound in a complex with another atom.

\section{Results}

\begin{figure*}
    \centering
    \subfloat[]{
        \includegraphics[width=0.32\textwidth]{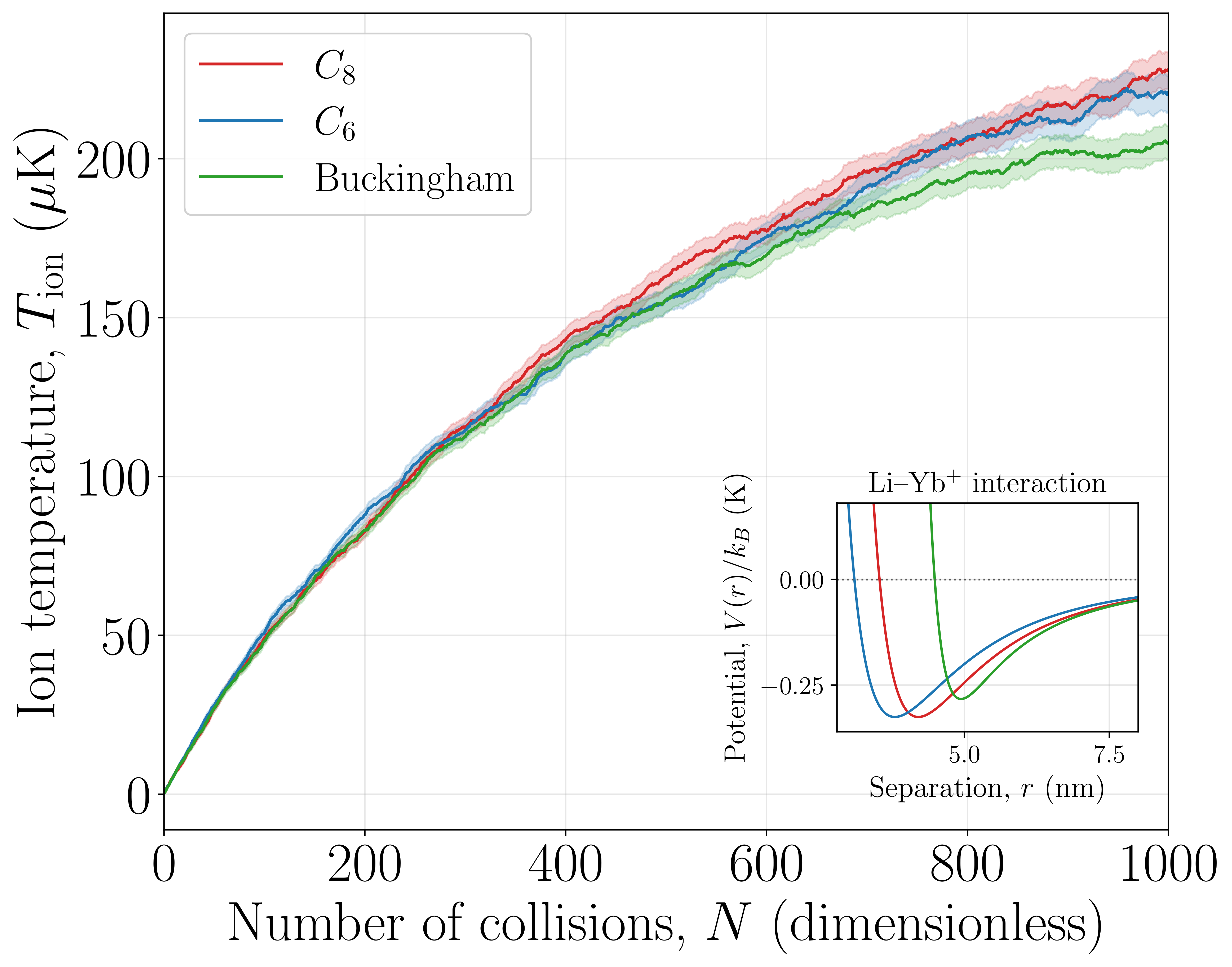}
    }
    \hfill
    \subfloat[]{
        \includegraphics[width=0.32\textwidth]{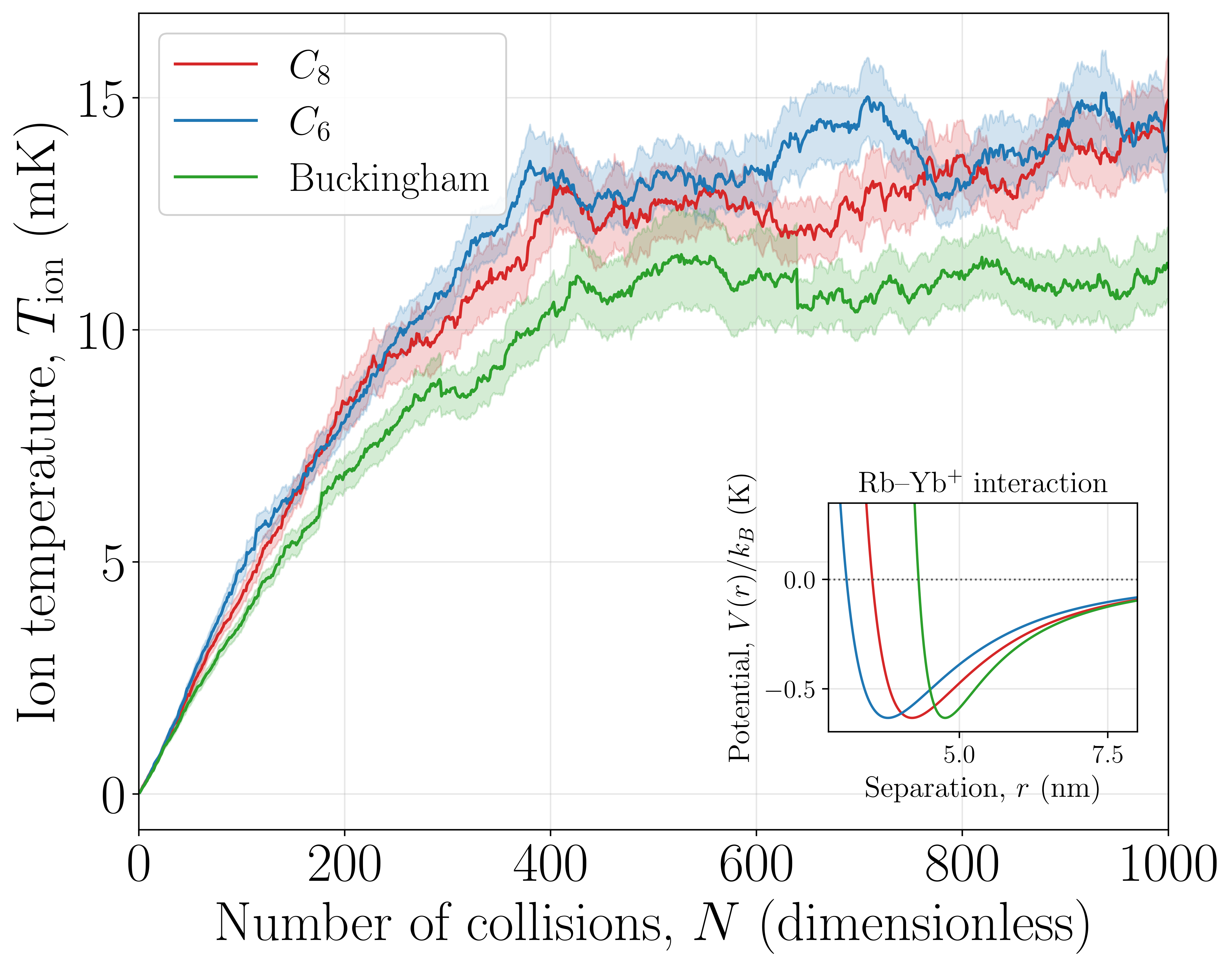}
    }
    \hfill
    \subfloat[]{
        \includegraphics[width=0.32\textwidth]{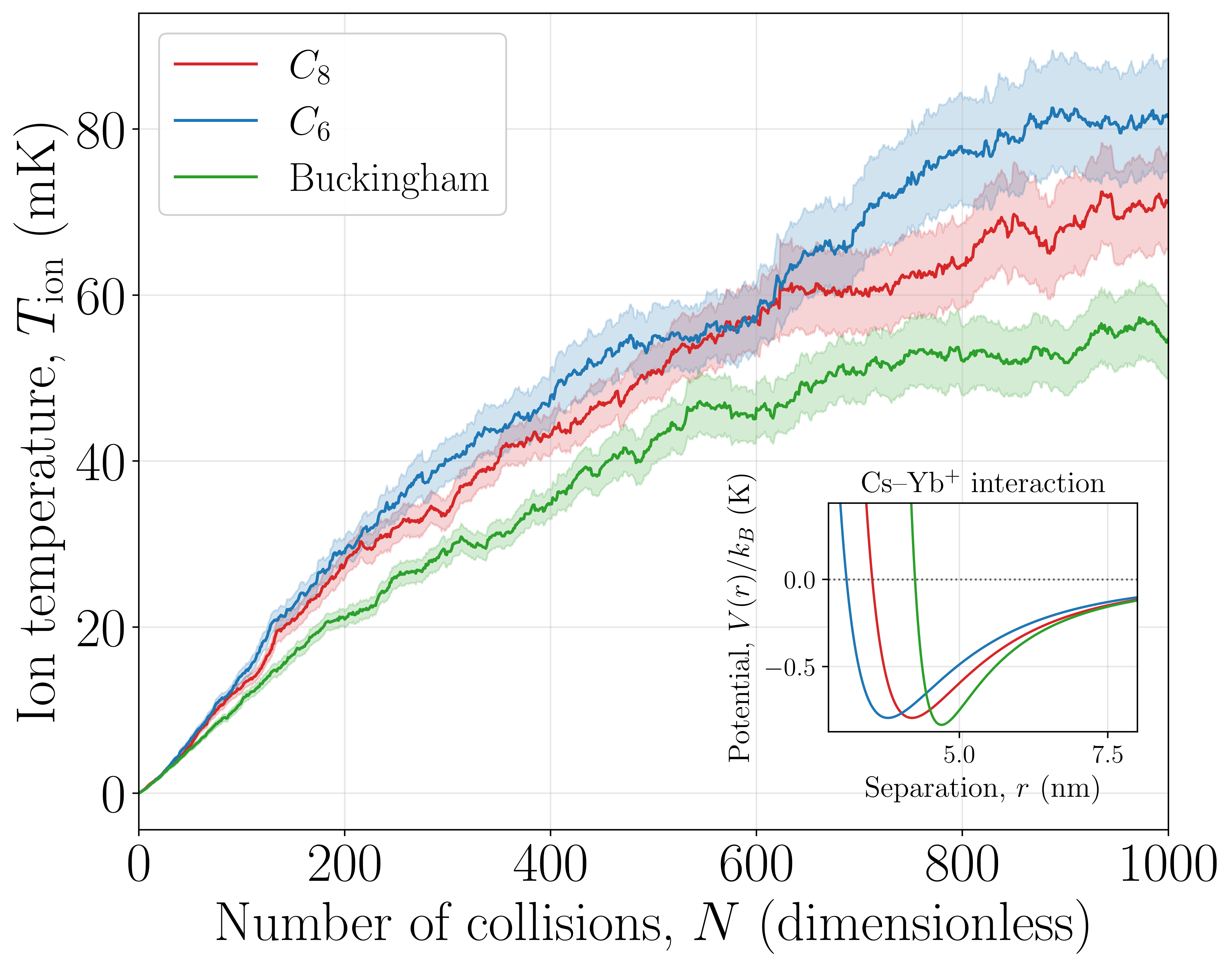}
    }
    \caption{Panels (a), (b), and (c) present the average temperature of the $^{174}$Yb$^+$ ion in a Paul trap ($\Omega_\textrm{rf}=2\pi\times 26.5$ MHz) as a function of the number of collisions with $^6$Li, $^{87}$Rb, and $^{133}$Cs atoms,
    respectively, computed with the one-atom-at-a-time approach. All potentials for the same atom and ion species share approximately the same depth. The atoms are initialized on the sphere whose radius
    is 529~nm ($1.6\times 10^{12}$ cm$^{-3}$). Different colors indicate
    potentials with different short-range forms. The shaded areas represent one standard deviation of the data set.}
    \label{fig:fig3}
\end{figure*}

First, we perform an exploration of three key ingredients on the cooling rate and thermalization via the one-atom-at-a-time approach: short-range effects and mass of the atomic bath. For this set of simulations we study a single $^{174}$Yb$^+$ ion in contact with a gas of lithium, rubidium and cesium. Each of this atomic gases is considered with the same denstiy of $10^{12}$ cm$^{-3}$, consistent with a sphere radius of 529 nm. For each of the three atomic species under consideration, we consider three distinct short-range atom-ion interaction potentials: $V_8(r)=C_8/r^8-C_4/r^4$, $V_6(r)=C_6/r^6-C_4/r^4$, and $V_B(r)=Ae^{-Br}-C_4/r^4$, where $C_4$ is fixed by the atom species considered. The results for the ion average kinetic energy as a function of the number of collisions is shown in Fig.~\ref{fig:fig3}, assuming a bath temperature of 2$\mu$K.

Heavier atoms systematically yield a larger average kinetic energy of the ion, consistent with previous observations and the unfavorable ion-to-atom mass ratio~\cite{Cetina2012,furst2018prospects,paul_trap1,Pinkas2020}. This is because among these three atoms the polarizability increases with the atom mass, inducing a more attractive long-range interaction and moving the ion more away from its initial position, at the center of the trap. In addition, because the 2-body center of mass is closer to the atom for heavier atoms, the larger attraction will cause the ion to have to move more away from the trap center to reach the center of mass before the collision. The further the ion is forced to move from the trap center per collision, the more micromotion-induced heating it will feel \cite{Cetina2012}, increasing its kinetic energy after many collisions. Interestingly, this heating effect is accompanied by greater sensitivity to the details of the short-range atom-ion interaction, reaching up to 20\% of difference in cesium, even though the interaction potentials share a very similar dissociation energy. 

\begin{figure}[h]
    \centering
    \includegraphics[width=\linewidth]{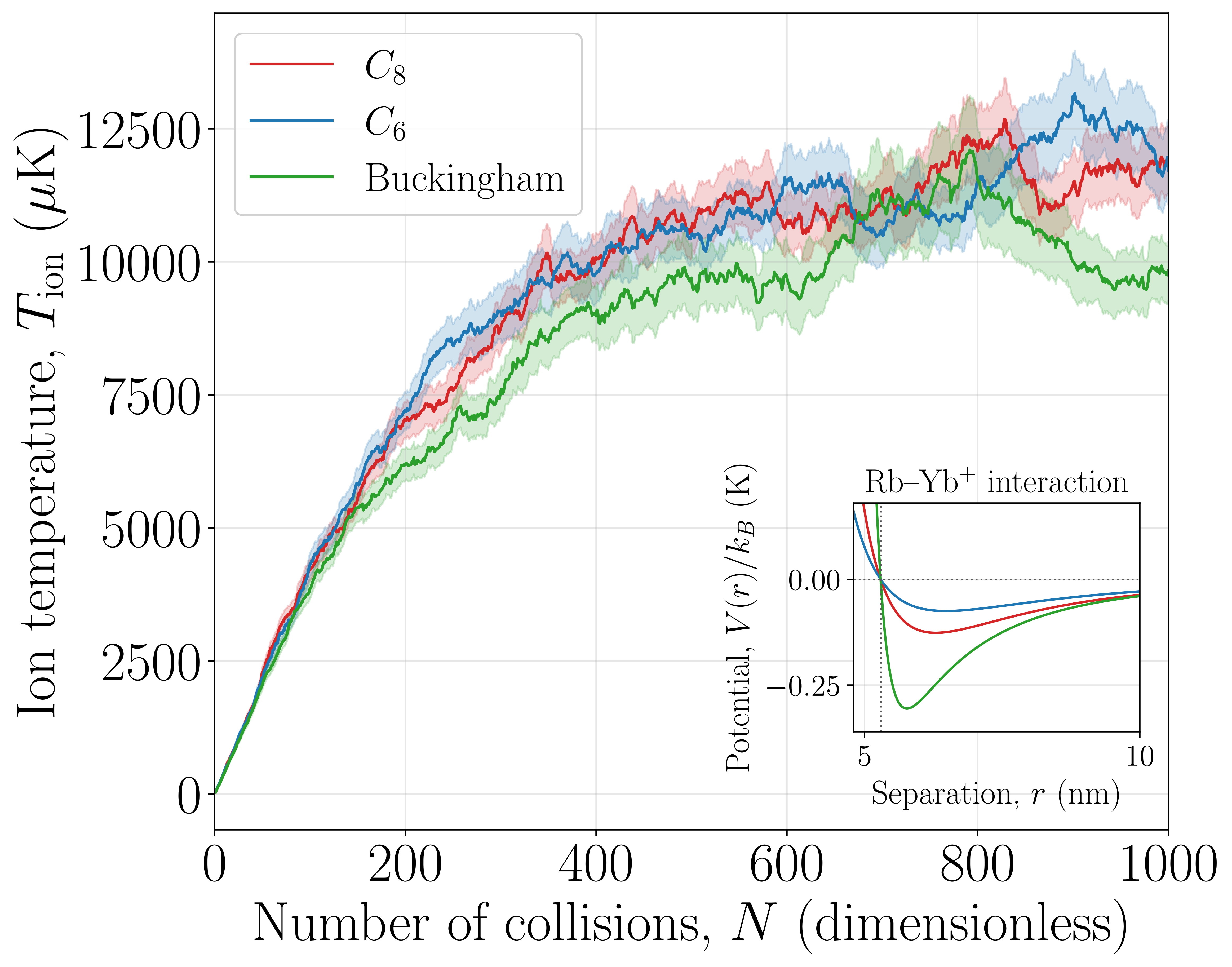}
    \caption{The average temperature of the $^{174}$Yb$^+$ ion as a function of the number of collisions with a $^{87}$Rb atom, which is on a 529~nm-radius sphere. Different colors indicate potentials with different short-range forms. The potentials share the same $r_0$ (5.29~nm). The shaded areas are one standard deviation.}
    \label{fig:fig4}
\end{figure}

To further explore the role of the short-range interaction, we compute the ion's average kinetic energy as a function of collision number in a Rb gas at $2\,\mu$K for several interaction potentials, shown in Fig.~\ref{fig:fig4}. Here, all potentials have different well depths, but share the same classical turning point at zero collision energy, $r_0$, which can be thought of as the barrier width with $V(r_0)=0$. By contrast, the potentials used in panel (b) of Fig.~\ref{fig:fig3} have the same well depth but different values of $r_0$. Under otherwise identical physical conditions, the potentials considered here yield nearly identical thermalization rates and steady-state average kinetic energies, unlike those in Fig.~\ref{fig:fig3}(b). These results support an interpretation based on the collision-induced displacement of the ion from the trap center. During each collision, the attractive interaction pulls the ion toward the atom before short-range repulsion reverses their relative motion. Over the first few hundred collisions, these repeated perturbations can increase the ion's typical displacement from the trap center. A smaller $r_0$ allows the particles to approach more closely and, in this picture, prolongs the attractive interaction. The resulting larger displacements expose the ion to stronger micromotion heating, increasing its kinetic energy until a steady state is established under secular confinement.

The dependence on $r_0$ is expected to be more pronounced for heavier atoms: for a given change in the relative separation, the ion accounts for a larger fraction of the displacement about the center of mass. Thus, the similar thermalization behavior obtained for potentials with a common $r_0$ supports our interpretation that the distance of closest approach, rather than the well depth, primarily controls the heating under the conditions considered. A larger $r_0$ limits the inward approach and the duration of the attractive interaction, reducing the ion's displacement from the trap center and yielding less micromotion heating and a lower steady-state kinetic energy.

To verify this mechanism directly, we performed an additional set of simulations consisting of 1000 independent head-on collisions between a $^{174}$Yb$^+$ ion and a $^{87}$Rb atom. In each simulation, the atom was initially placed 529~nm from the ion, while its initial velocity was randomly sampled from the thermal distribution of the atomic gas. Five datasets were generated by varying $C_8$ to test different short range interaction potentials with $r_0$ equal to 2.7, 4.0, 5.3, 6.6, and 7.9 nm. We find that the average displacement of the ion from the trap center decreases monotonically from 29.4~nm to 28.9~nm as $r_0$ is increased from 2.7~nm to 7.9~nm. This provides direct evidence that the distance of closest approach controls the micromotion-induced heating of the ion after one collision. This result demonstrates that the potential barrier width is a key parameter governing the thermalization dynamics of trapped ions in ultracold atomic gases. More generally, it suggests that $r_0$, despite often being overlooked, provides a more physically meaningful descriptor of the short-range interaction than the potential depth alone.

\subsection{Comparison between full MD simulations and one-atom-at-a-time}

\begin{figure*}
    \centering
    \includegraphics[width=\linewidth]{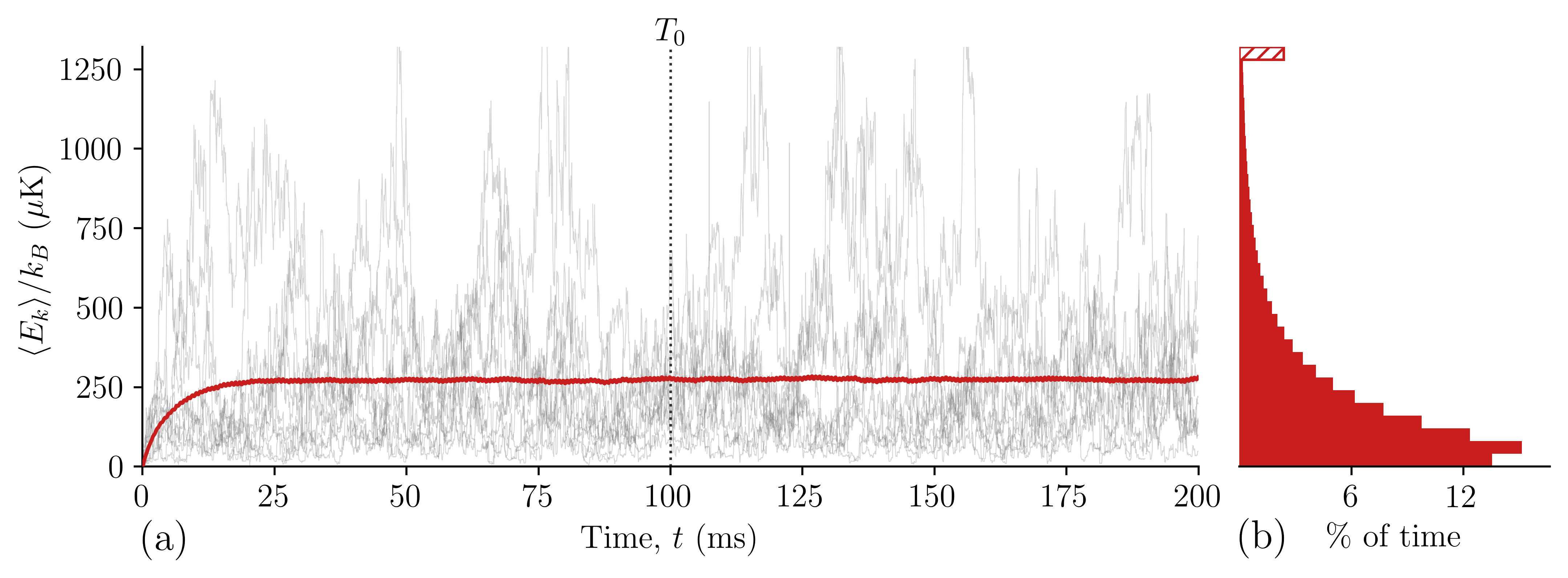}
    \caption{Thermalization of a single trapped $^{174}$Yb$^+$ ion in an ultracold $^7$Li buffer gas. The ion is initialized at rest in the center of a Paul trap ($\Omega_\textrm{rf}=2\pi\times 26.5$ MHz) and the gas is held at 2 $\mu$K. The atom-ion interaction is $V(r) = -C_4/r^4 + C_8/r^8$ with $C_4=82.056$ a.u., potential minimum $R_e = 4.205$ nm and well depth $D_e/k_B = 325.1$ mK; the gas density is $1.61\times 10^{12}$ cm$^{-3}$. The ion heats from rest within about 25 ms and then fluctuates about a steady state of $\langle E_k\rangle/k_B = 273\pm 4$ $\mu$K, two orders of magnitude greater than the buffer-gas temperature. In Panel (a), the curve is the mean ion kinetic energy, $\langle E_k\rangle/k_B$, calculated over up to 8560 independent trajectories with different random seeds, during the first 200 ms of simulated time (with the fixed number of timesteps, only 6561 seeds remain at 100ms, and only 1484 seeds remain at 200ms). Each data point is averaged over about 50 $\mu$s. Panel (b) shows the distribution of ion instantaneous kinetic energies during the 6561 trajectories after $T_0=100$ ms whenever the ion is not in a complex, and the average of these energies yields the ion's average kinetic energy which is one of the points plotted in Fig.~\ref{fig:fig6}.}
    \label{fig:fig5}
\end{figure*}

In order to test the accuracy of the one-atom-at-a-time approach with Weibull-distributed atom velocity radial components and Gaussian velocity tangential components whose results were studied in detail in the previous section, we also run a ``full'' molecular dynamics simulation of the trapped ion in a buffer gas. In this full MD simulation, there are always 10 atoms within a sphere surrounding the ion, such that whenever an atom leaves, a new atom emerges through the sphere's surface with velocity components sampled from Gaussians at the bath temperature, but only accepted to spawn with a probability proportional to the radial component of the relative velocity of the proposed atom with respect to the ion, as described in Section~\ref{sampling}. All atom-ion interactions are included, but atom-atom interactions are not \footnote{See [27].}

The results of a full MD simulation of $^{174}$Yb$^+$-$^7$Li during the first 200~ms are shown in Fig.~\ref{fig:fig5}. The density in this case is $1.61\times 10^{12}$ cm$^{-3}$, with the lithium atoms at 2~$\mu$K. We run thousands of independent instances of a simulation of a trapped ion interacting with 10 atoms, such that whenever an atom drifts too far from the ion, another atom is spawned in, maintaining constant average density and bath temperature, thus emulating the scenario of the ion in a uniform neutral gas. Each instance is ran for a fixed number of adaptive timesteps, which we choose by estimating at what time the ion will be thermalized, $T_0$, and then estimating the number of timesteps it would take for a trajectory without a molecular ion to reach $2T_0$; for the system in Fig.~\ref{fig:fig5}, $T_0=100$ ms and we estimate 350 million timesteps, which is the same order of magnitude as the number of timesteps it takes the one-atom-at-a-time simulation in Fig.~\ref{fig:fig3}(a) to thermalize. The ion starts with zero kinetic energy at the center of the trap, and is given sufficient time to thermalize with the surrounding atoms. What we find is that the ion thermalizes to an average kinetic energy of $273\pm 4$~$\mu$K. This is 2 orders of magnitude greater than the bath temperature because the atoms pull the ion away from the trap center, inducing micromotion heating from the trap's rf fields. Also, the ion forms many trap-assisted complexes with the atoms, which temporarily increase the temperature during the run, as can be seen in the faint lines in Fig.~\ref{fig:fig5}. These contribute to the long tail of the distribution of the ion's average kinetic energy after 100 ms, shown in Fig.~\ref{fig:fig5}(b). We find that in many of the simulations, the ion eventually ends up in a very-long lived complex (tens of milliseconds). While in this complex, the ion vibrates very rapidly, causing it to obtain a simulated average kinetic energy orders of magnitude beyond $200$ $\mu$K. The nature of these especially long-lived, non-chaotic complexes, that resemble molecular ions, will be described in a separate study. For now, in the reported temperature we exclude all timesteps during which the ion is in a trap-assisted complex or molecular ion. 

\begin{figure}
    \centering
    \includegraphics[width=\linewidth]{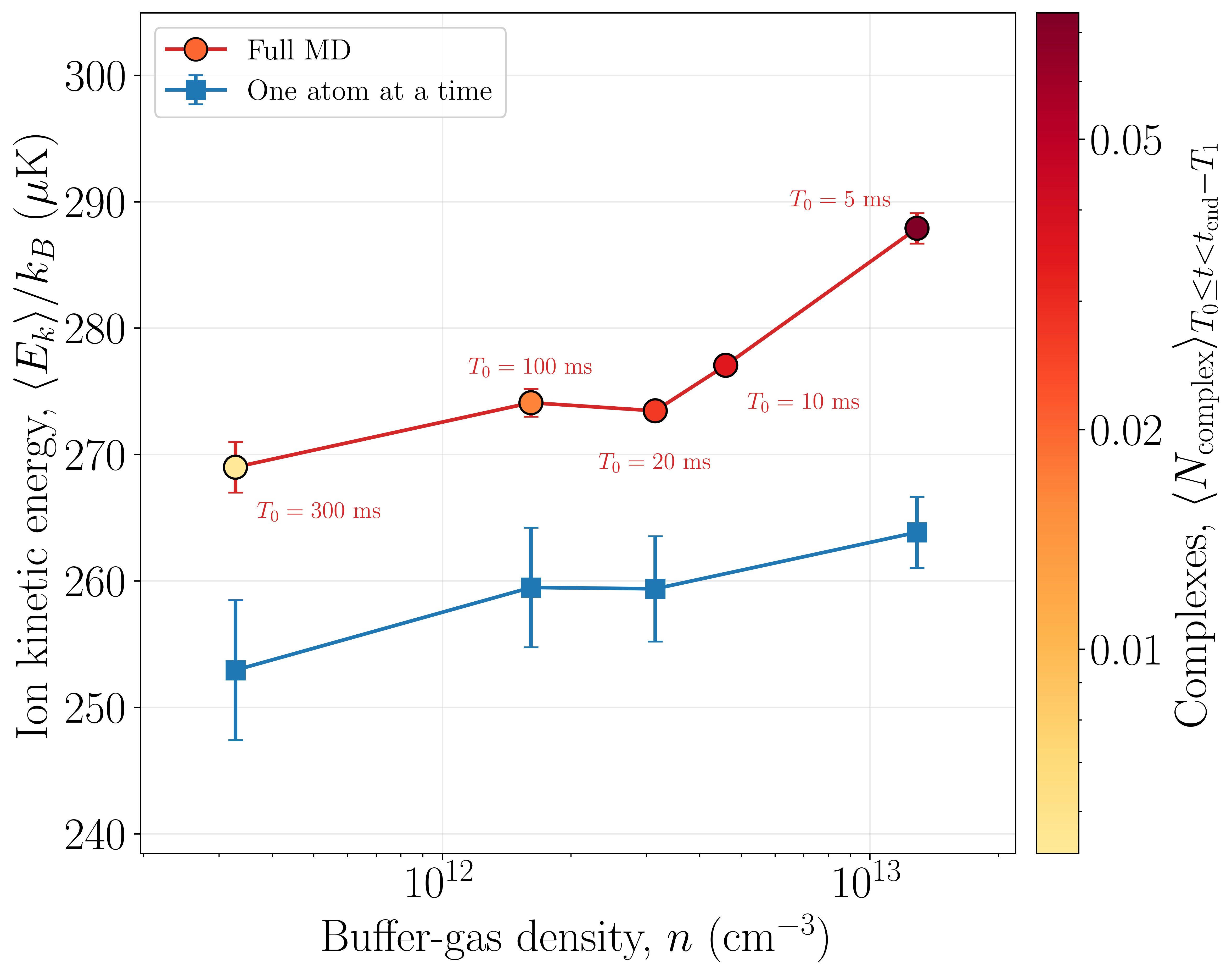}
    \caption{Comparison of one-at-a-time and full MD approaches as a function of density for the $^{174}$Yb$^+$-$^7$Li system with $\Omega_\textrm{rf}=2\pi\times 26.5$ MHz. Here, $T_0$ is the thermalization time, $T_1$ is the maximum amount of time spent in the final molecular ion, and $t_\textrm{end}$ is the total simulation time for a given seed. The color indicates the average number of atoms in a quasi-bound state with the ion, averaged after $T_0$ but before molecular ions start to form at $t_\textrm{end}-T_1$. The one-at-a-time simulations (1023 trajectories total) for the three higher densities ran for 4000 collisions and we calculated the average over the last 2000 collisions; the lowest density was run for 8000 collisions and then we report the value of $A$ in a fit to $E(t)=A(1-\exp(-t/\tau))$ ($R^2>0.99$) where $t$ is the number of collisions.}
    \label{fig:fig6}
\end{figure}

\subsection{Atomic bath density effects}

Fig.~\ref{fig:fig5} is just one single physical ensemble (one trap parameter set, one short range, one atom species, one density). In this work, we use full MD to study a total of 26 ensembles. For each ensemble, we calculate the ion's average kinetic energy, found by averaging over all recorded instantaneous kinetic energies after the thermalization time $T_0$ when the ion wasn't in a complex, over all seeds of the ensemble. The ion's average kinetic energies for 4 of these ensembles are shown in Fig.~\ref{fig:fig6}. These 4 ensembles only differ in density, which amounts to changing the sphere radius for spawning new atoms. We run both the one-at-a-time approach in blue and the full MD (10-atom) approach. First, we notice that the full MD results systematically lead to slightly larger average kinetic energies of the ion, with an increment of $\lesssim 10$\%. 
The one-at-a-time approach does not show a dependence on the density, as expected based on previous simulations~\cite{Feldker2020}, although the lower-density runs take much longer to converge due to the increase in sphere radius making it quadratically less likely for an atom spawned on the sphere surface to actually collide with the ion.  The full MD approach, on the other hand, yields average kinetic energies of the ion which increase with density at large densities. This correlates with the average number of complexes the ion is in during the run after thermalizing and before forming a molecular ion, as shown by the colorbar. What this means is that if there is a greater chance for a trapped ion in a buffer gas to be found in a complex at any given time, then the ion's measurable temperature whenever it is not in a complex will be higher, in agreement with previous predictions in the case of the one-atom-at-a-time approach~\cite{Chowdhury_2025}.

This behavior can be attributed to two mechanisms. First, the ion gains kinetic energy upon dissociation of an atom--ion complex, so more frequent complex formation and dissociation lead to greater heating. Second, long-lived complexes can undergo collisions with a third atom, which may quench the complex and produce additional ion heating. These effects are weaker at lower bath densities: the rates of complex formation and collisions with bath atoms scale linearly with density, whereas the rate of three-body events scales quadratically. Nevertheless, the density dependence remains modest over the range considered, accounting for only 5\%--10\% of the ion's temperature.

\begin{figure}
    \centering
    \includegraphics[width=\linewidth]{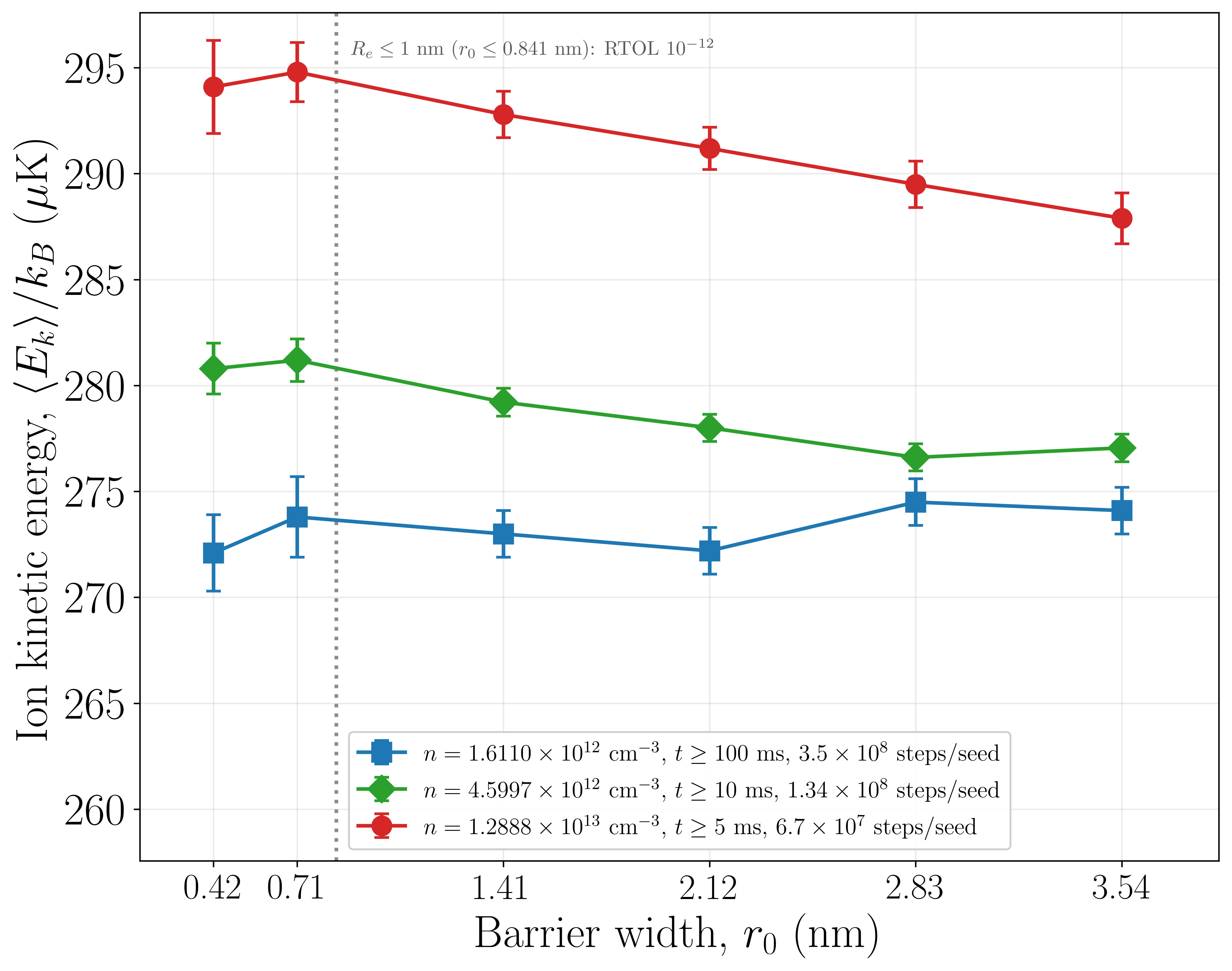}
    \caption{Dependence of the ion's average kinetic energy on the short-range physics, for three different densities, for a $^{174}$Yb$^+$ ion in a trap with $\Omega_\textrm{rf}=2\pi\times 26.5$ MHz and 2$\mu$K bath of $^7$Li atoms. The model potential is $V(r)=C_8/r^8-C_4/r^4$ with $C_4$ fixed by lithium, and we vary $C_8$ to obtain potentials with the same long range but equilibrium distances ranging from 0.5nm - 4.2nm. The relative and absolute tolerances were $(10^{-12},10^{-24})$ for $r_0<0.8$nm and $(10^{-10},10^{-20})$ for $r_0>0.8$nm.} 
    \label{fig:fig7}
\end{figure}

\subsection{Short-range atom-ion interaction effects}

Having established the consistency between the full-MD approach and the one-at-a-time approach at low densities, and the ability of the full-MD approach to capture minute density effects, we proceed to use the full-MD approach to explore the effect of the short-range potential barrier width $r_0$ on the ion's average kinetic energy before molecular ions form, at 3 different densities, shown in Fig.~\ref{fig:fig7}. First of all, just as we found that the density yields only a 5\%-10\% effect on the ion's temperature (which Fig.~\ref{fig:fig7} shows holds regardless of the short-range physics), we further find that atom-ion $r_0$ yields an even smaller effect, less than 5\% on the ion's average kinetic energy. This effect is more noticeable at higher densities, meaning that the ion's kinetic energy is more sensitive to the short range at higher buffer gas densities, since the interparticle spacing for densely packed atoms is more likely to be near short-range. 

So, for higher densities, the ion's average kinetic energy decreases by about 5\% as the classical turning point at zero collision energy increases from 0.5~nm (a realistic number) to 4~nm (closer to typical simulation model potentials). This small effect can be attributed to the short-range encounters which don't form complexes, when the ion moves significantly faster as it travels between 4~nm to 0.5~nm away from atoms. At the shortest ranges, the small effect disappears because the shorter time spent at the closest range overpowers the effect of getting faster. 

\begin{figure}[h]
    \centering
    \includegraphics[width=\linewidth]{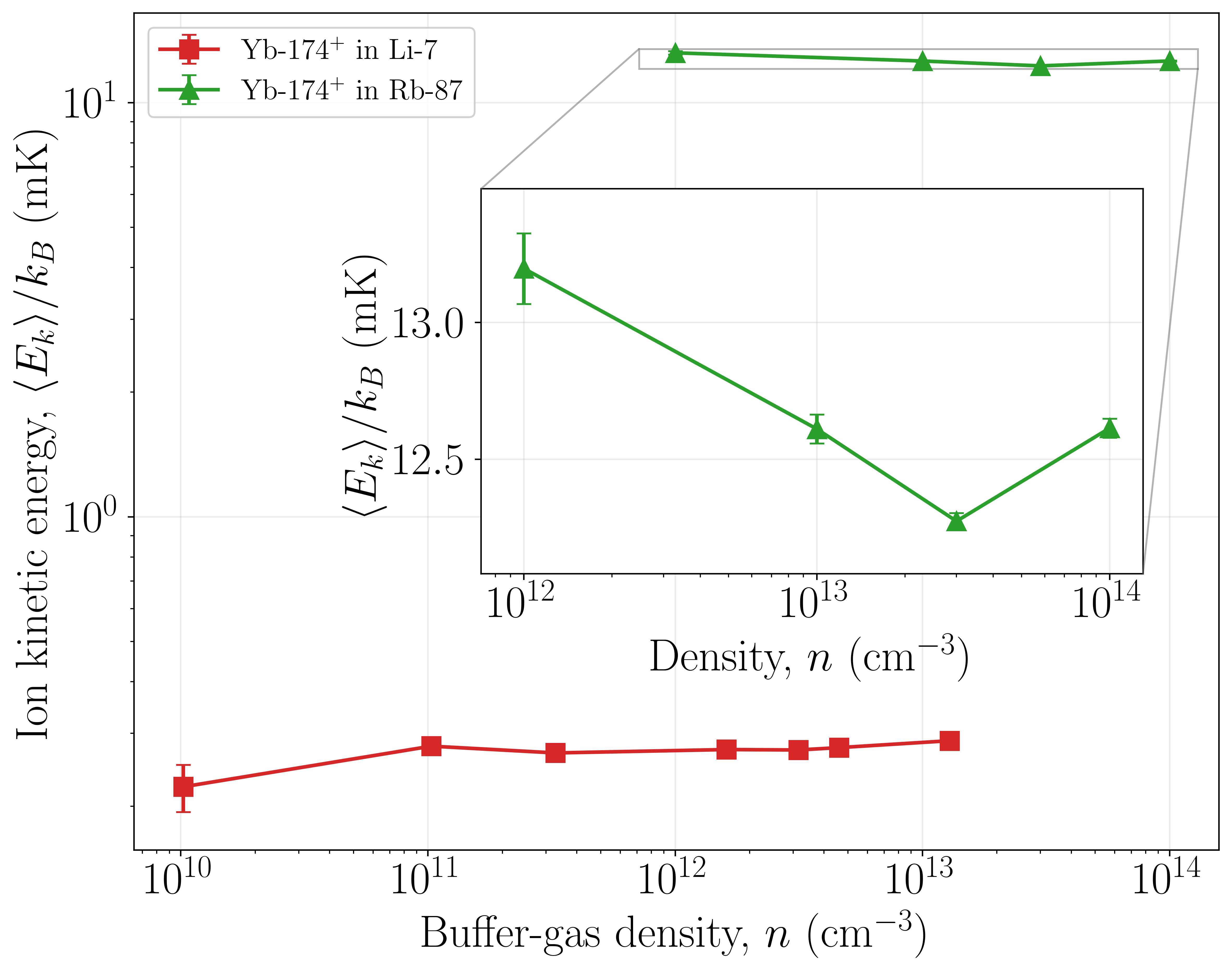}
    \caption{The ion's average kinetic energy versus atom bath density, for an experimental trap setup \cite{Pinkas2020} with $\Omega_\textrm{rf}=2\pi\times 26.5$ MHz, substituting the system with $^{174}$Yb$^+$-$^7$Li and $^{174}$Yb$^+$-$^{87}$Rb.}
    \label{fig:fig8}
\end{figure}

\subsection{The role of the atom mass}

Finally, we apply our full MD approach to another experimentally accessible system with the same physical parameters as in Fig.~\ref{fig:fig6}, except that we replace the atoms with $^{87}$Rb. Fig.~\ref{fig:fig8} shows our approach applied to a $^{174}$Yb$^+$ ion in a bath of $^{87}$Rb atoms, with the same trap parameters as those we computed for the $^{174}$Yb$^+$ ion in a $^{7}$Li bath, also shown in the figure for comparison. Comparing Rb to Li shows that atomic mass has by far the greatest effect on temperature of all the parameters we studied so far using full MD simulations, fully consistent with the one-at-a-time simulation results in Fig.~\ref{fig:fig3}(a,b). In the inset, which shows a zoom-in of the Rb atomic bath, we see that the ion's average kinetic energy varies by less than about 5\%, indicating that density may be less relevant for heavier atomic baths.

\section{Conclusion}

We have developed a full molecular dynamics simulation approach to simulate the sympathetic cooling of a single trapped ion immersed in an ultracold atomic gas. The method uses the ion's velocity to define an effective interaction region in which a controlled number of atoms interact with the ion throughout the simulation, allowing repeated scattering and simultaneous interactions with multiple atoms. This framework enables us to examine density-dependent effects beyond those captured by one-atom-at-a-time simulations. Over the range of densities considered, density-related effects contribute no more than approximately 10\% to the ion's steady-state average kinetic energy. We attribute this contribution to the formation of trap-assisted atom--ion complexes and their subsequent stabilization through collisions with additional atoms, processes absent from the one-atom-at-a-time approach. Including multiparticle dynamics yields systematically higher steady-state average kinetic energies than the corresponding one-atom-at-a-time simulations, indicating that the latter underestimate ion heating under the conditions studied. At higher densities, the smaller mean interatomic separation increases the frequency of encounters involving the short-range region of the interaction potential. Nevertheless, the resulting sensitivity of the steady-state average kinetic energy to the short-range potential remains below 3\%. We also find a strong dependence on the atomic mass, consistent with previous studies based on one-atom-at-a-time simulations.

The full molecular dynamics simulation approach provides a more complete description of ion thermalization in a neutral gas and a benchmark for assessing the limitations of simulations restricted to successive two-body encounters. It also offers a framework for investigating the formation of more complex reaction products beyond trap-assisted atom--ion complexes. The implementation can be extended to single-ion dynamics in multipole Paul traps and to multiple ions interacting with a buffer gas, although the latter extension would require substantially greater computational resources.

\begin{acknowledgements}
    This work was supported by the U.S. National Science Foundation through CAREER Award No. PHY-2440808. 
\end{acknowledgements}

\section*{Data and code availability}

The data that support the findings of this article and the simulation code used to generate them are available from the corresponding author upon reasonable request.

\bibliographystyle{unsrt}

\end{document}